\documentclass[final,5p,times,twocolumn]{elsarticle}

\usepackage{graphicx}
\usepackage{epstopdf}
\journal{Physics Letters B}
\begin{document}

\begin{frontmatter}

\title{Oscillations above the barrier in the fusion of $^{28}$Si + $^{28}$Si}

\author{G.~Montagnoli\corref{*}$^1$, A.~M.~Stefanini$^2$, H.~Esbensen$^3$,  L.~Corradi$^2$, S.~Courtin$^4$, E.~Fioretto$^2$, 
J.~Grebosz$^{5}$, F.~Haas$^4$, H.~M.~Jia$^{2}$, C.~L.~Jiang$^3$, M.~Mazzocco$^{1}$, C.~Michelagnoli$^{1}$, T.~Mijatovi\'{c}$^6$, D.~Montanari$^1$, C.~Parascandolo$^{1}$, F.~Scarlassara$^{1}$, E.~Strano$^{1}$, S.~Szilner$^6$, D.~Torresi$^{1}$ \\}

\address{\medskip
$^1$ Dipartimento di Fisica e Astronomia, Universit\`a di Padova, and INFN, Sez. di Padova, I-35131 Padova, Italy\\
$^2$ INFN, Laboratori Nazionali di Legnaro, I-35020 Legnaro (Padova), Italy\\
$^3$ Physics Division, Argonne National Laboratory, Argonne, IL 60439, USA\\
$^4$ IPHC, CNRS-IN2P3, Universit\'e de Strasbourg, F-67037 Strasbourg Cedex 2, France\\
$^5$ Institute of Nuclear Physics, Polish Academy of Sciences, PL 31-342 Cracow, Poland\\
$^6$ Ru{d\llap{\raise 1.22ex\hbox
      {\vrule height 0.09ex width 0.2em}}\rlap{\raise 1.22ex\hbox
      {\vrule height 0.09ex width 0.06em}}}er
      Bo\v{s}kovi\'{c} Institute, HR-10002 Zagreb, Croatia
      
 \cortext[*]{Corresponding author. \\{\it E-mail address:} montagnoli@pd.infn.it }
}

\date{\today}

\begin{abstract}

Fusion cross sections  of $^{28}$Si + $^{28}$Si have been measured in a range above the barrier with a very small energy step  ($\Delta$E$_{lab}$ = 0.5 MeV). Regular oscillations have been observed, best evidenced in the first derivative of the energy-weighted excitation function.  For the first time, quite different behaviors (the appearance of oscillations and the trend of sub-barrier cross sections) have been reproduced within the same theoretical frame, i.e., the coupled-channel model using the shallow M3Y+repulsion potential.
The calculations suggest that channel couplings play an important role in  the appearance of the oscillations, and that the simple relation between a peak in the derivative of the energy-weighted cross section and the height of a centrifugal barrier is lost, and so is  the interpretation of the second derivative of the excitation function as a barrier distribution for this system, at energies above the Coulomb barrier.

\end{abstract}


\end{frontmatter}
\section{Introduction}
\label{intro}

Heavy-ion fusion dynamics near and below the Coulomb barrier is a matter of continuing interest, since it allows a deep insight into the fundamental problem of quantum tunnelling of many-body systems facilitated by channel coupling effects. Cross section enhancements, barrier distributions and, more recently, fusion hindrance effects have been observed and are being investigated.
\par
 Moreover, oscillatory structures were evidenced  a long time ago in the fusion excitation function of light heavy-ion systems like $^{12}$C + $^{12}$C, $^{12}$C + $^{16}$O and $^{16}$O + $^{16}$O~\cite{Sperr,Sperr2,Kovar,Tserruya}, in the energy region above the Coulomb barrier. Analogous oscillations were found for $^{20}$Ne + $^{20}$Ne~\cite{Poffe}. In that work, it was suggested for the first time that such features are due to successive partial waves entering the fusion cross
section as their centrifugal barriers are exceeded. In such cases the separation between nearby barriers is large with respect to the intrinsic energy width associated with the quantal penetration, so that the oscillations may become observable.
\par
This topic was further analyzed in the more recent work of Esbensen~\cite{Esbeosc}, where earlier data on $^{28}$Si + $^{28}$Si~\cite{Gary} were compared to detailed calculations in the coupled-channels (CC) model, along with the experimental evidences on lighter systems. 
Those previous data~\cite{Gary} on $^{28}$Si + $^{28}$Si above the barrier have experimental errors and energy steps just too large to allow a clear-cut conclusion about the existence of oscillating structures. 
A more detailed investigation for this system was performed by Aguilera at al. in Ref.~\cite{Agui} using the $\gamma$-ray technique, where the authors did not observe any oscillation in the range E$_{lab}\simeq$58-100 MeV.

\par
The existence of such structures, and the detailed effects produced by channel couplings in the energy
dependence of the fusion cross sections, can be best revealed by the
first and second derivatives of the excitation function multiplied by the energy (the so-called ``energy-weighted excitation function'')

\begin{equation}
D(E) = \frac{d(E\sigma_f)}{dE} , \hskip 1cm B(E) = \frac{d^2(E \sigma_f)}{dE^2} ,
\label{cent}
\end{equation}

\noindent where $E$ is the center-of-mass energy. As discussed in detail in Ref.~\cite{Esbeosc}, 
and following the Hill-Wheeler expression, each partial-wave cross section behaves like a Fermi function in the no-coupling limit, so that $D(E)$ is the sum of individual centrifugal barriers weighted with the factor (2$L$ +
1), each one centered at $V_b$(L). 
\par

The concept of a barrier distribution as defined by B(E) has been very helpful in the analysis of several sets of fusion data near and below the Coulomb barrier for medium-mass and heavy systems, often giving a "fingerprint" of the relevant coupled channels.
Its definition was inspired by Wong's formula whose second derivative is a symmetric distribution centered at the s-wave Coulomb barrier. 
\par
The barrier distribution $B(E)$ does not carry any information
about the individual $L$-dependent barriers. Indeed, when the peaks of neighboring $L$ values overlap strongly, one can sum up their contributions as in Wong's formula, and the barrier distribution is then given by B(E). The overlap
between near-by peaks, however, diminishes with increasing $L$, which typically results in the
breakdown of Wong's formula above the barrier (and of the interpretation of B(E) as a barrier distribution). This has been recognized~\cite{Esbeosc} in light, symmetric systems like $^{16}$O + $^{16}$O where only even values of L contribute to fusion and the peaks are more distant from each other, but not in the fusion of heavy systems, where the condition for separating the individual centrifugal barriers (Eq.(11) of Ref.~\cite{Esbeosc}) requires large $L$ values where many reaction channels open up and smear out the structures. 

The hindrance phenomenon of heavy-ion fusion  
at deep sub-barrier energies has been observed for several systems~\cite{Jiang0,Jiang2,DasguptaPRL,NiFelast} in the last decade. 
The onset of fusion hindrance has often been associated with the energy where the logarithmic derivative, 
 \begin{equation}
  L(E) = \frac{1}{E\sigma_f} \frac{d(E\sigma_f)}{dE},
  \end{equation}
 reaches the value (named L$_{CS}$)
expected for a constant astrophysical $S$-factor~\cite{Jiang04}. 
At that energy the $S$-factor develops a maximum as a function of the energy.
However in several cases the hindrance effect is 
 not strong enough to produce an $S$-factor maximum~\cite{Back}. 
\par
The phenomenon is very intriguing and far from being fully understood. The structure of the two colliding nuclei~\cite{4040} and, possibly, couplings to transfer channels~\cite{PLB,HMJia}, affect the energy threshold below which hindrance shows up.
\par
The M3Y+repulsion potential of Ref.~\cite{Misicu} produces a relatively shallow
  potential in the entrance channel and it has been capable of explaining 
  the fusion hindrance phenomenon in many cases, when applied in 
  CC calculations. 
  There are other models on the market that can explain the fusion
  hindrance phenomenon without employing a shallow potential, for
  example, the model by Ichikawa et al.~\cite{Ichi1}. In that model
  the hindrance is caused by the damping of collective excitations for
  overlapping nuclei~\cite{Ichi2}. In order to resolve the differences
  between the two models, it is therefore of great interest to test 
  these models against new observables, such as the oscillations that 
  appear in the measured cross sections at high energy.
\par
Hindrance effects are more clearly observed in heavier systems, where, on the other hand, possible consequences of stronger channel couplings also deserve attention.
An intermediate case like $^{28}$Si + $^{28}$Si calls for interesting investigations. High-precision data in a sufficiently wide energy range would allow exploiting (and checking) both definitions of Eq.(\ref{cent}) to compare the experimental results with detailed CC calculations. 
\par
The purpose of this work is then twofold: 1) to search for oscillations in the fusion excitation function for $^{28}$Si +$^{28}$Si  above the barrier by careful measurements with very small energy steps, and 2) to obtain a consistent interpretation of both the sub-barrier fusion excitation function~\cite{2828_2014} and of the oscillations within the same theoretical CC model.
A partial and preliminary account of the experimental part of this work was given in Ref.~\cite{INPC_2013}.

\section{Experimental}

$^{28}$Si beams  with intensities $\simeq$15-30 pnA
were delivered by the XTU Tandem accelerator of the Laboratori Nazionali di Legnaro of INFN.  Targets of  $^{28}$Si (with an isotopic enrichment of  99.93$\%$) 50 $\mu$g/cm$^2$  thick  on 15 $\mu$g/cm$^2$ carbon backings facing the beam, were used. 
\par
Two separate series of careful measurements of the excitation function (named I and II run) have been performed in the energy range $\simeq$62-78 MeV (above the barrier), with a step
small enough ($\Delta$E$_{lab}$ = 0.5 MeV) to resolve possible oscillations.
For each energy, at least 10000 fusion evaporation residues (ER) were detected, thus reducing the statistical error to 1$\%$ or less. 
The relevance of the accuracy of the $^{28}$Si beam energy in such measurement is obvious.
A particular care has been devoted to this issue, by stepping the field in the 90$^{\rm o}$ analysing magnet of the accelerator only downwards, so to minimise possible hysteresis effects. The maximum uncertainty in the beam energy was measured to be $\pm$0.13$\%$~\cite{stefanini95}   ($\simeq$90 keV at 70 MeV).  When the energy is monotonically decreased, as in the present experiments, the relative beam energy uncertainty is a factor 3-4 lower.
\par
The ER were detected at $\theta_{lab}$= 3$^{\rm o}$ using the same set-up and procedures described in Refs.~\cite{NiFelast,4040}, based on an electrostatic beam separator. This set-up is very simple to operate, allowing fast and reliable measurements of relative and absolute cross sections. In the present case, the absolute scale was fixed by normalising the relative yields to the cross sections of Ref.~\cite{2828_2014} at corresponding energies.

\begin{figure}
\begin{center}
\hspace*{-3mm}\includegraphics[width=8.53cm]{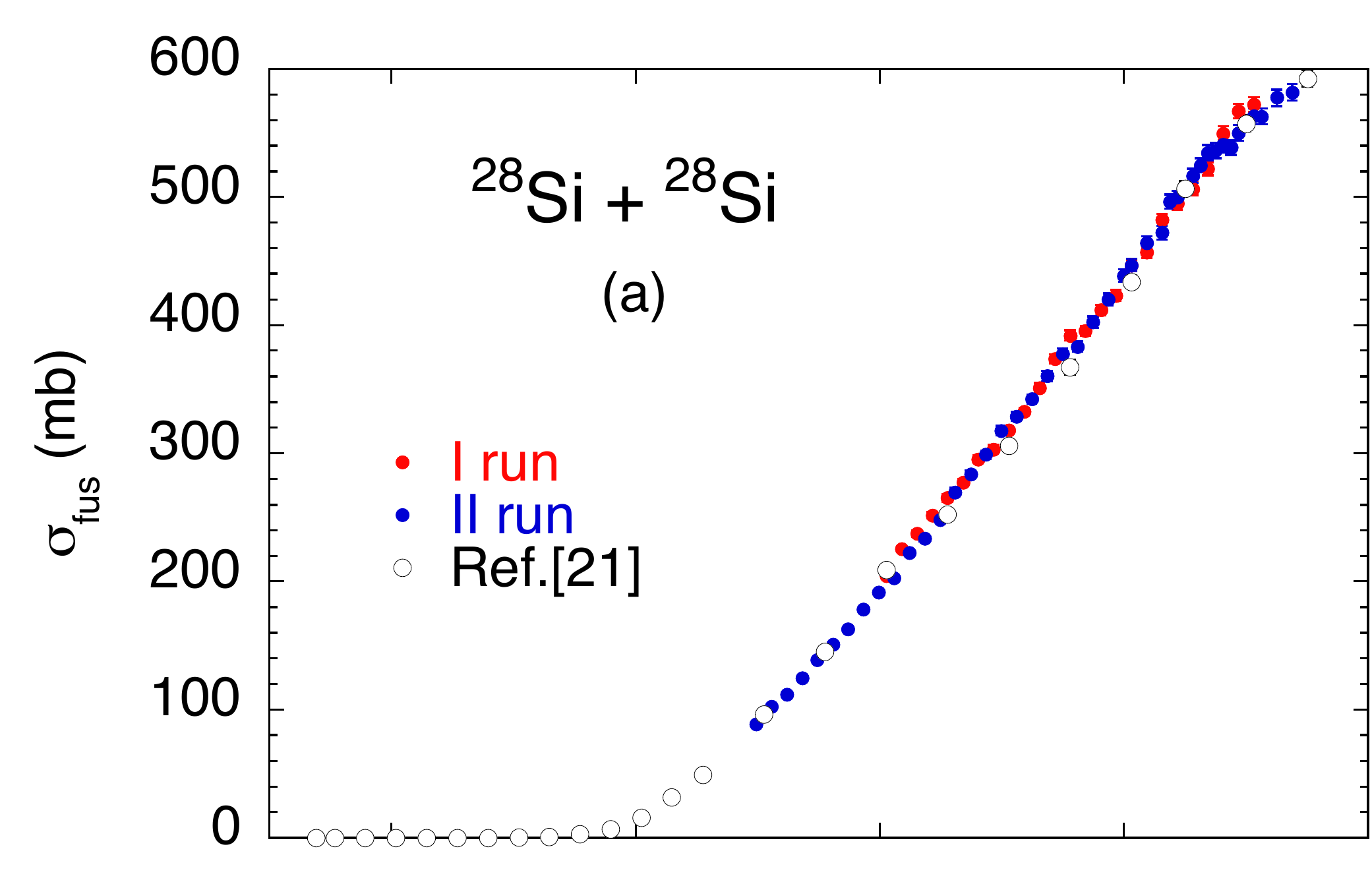}\vspace*{-4mm}
\includegraphics[width=9.15cm]{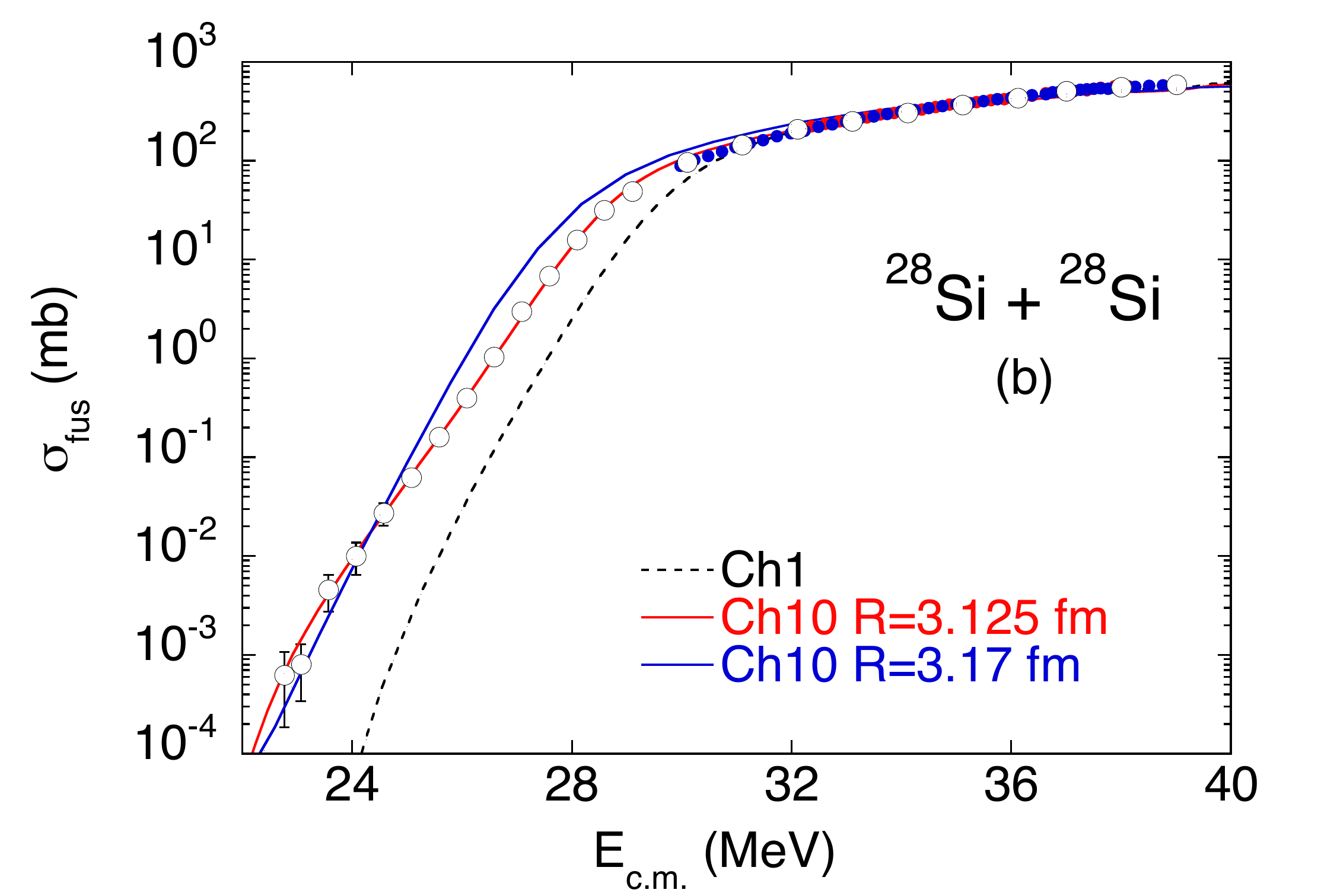}
\caption{(color online) Fusion excitation function of $^{28}$Si +$^{28}$Si in linear  
(a) and logarithmic  
(b) scales. 
The results of the CC calculations discussed in the text are also shown in (b).}
\label{fig1}
\end{center}
\end{figure}

\begin{figure}
\begin{center}
\includegraphics[width=8.5cm]{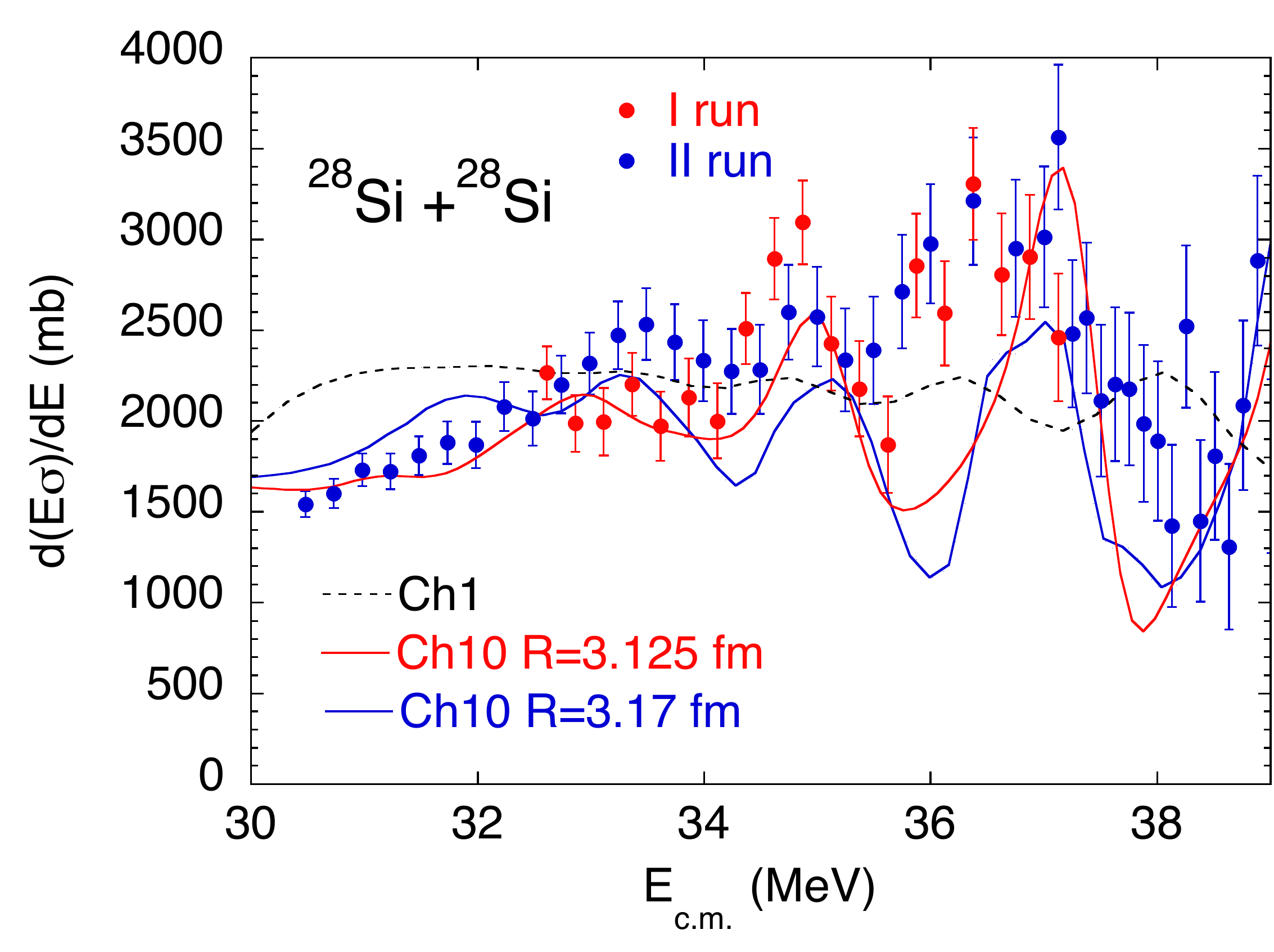}
\caption{(color online) The derivative of the energy-weighted  fusion excitation function D(E) of $^{28}$Si +$^{28}$Si in an energy range above the barrier. The derivative is obtained as the incremental ratio between successive points, with an energy step $\Delta$E$_{cm}$ = 0.75 MeV.
The  lines are the results of the theoretical calculations discussed in the text.}
\label{fig2}
\end{center}
\end{figure}

Fig.~\ref{fig1} shows the excitation function  of $^{28}$Si +$^{28}$Si 
in linear and logarithmic energy scales.
The excitation function above the barrier looks very smooth at first sight, but a closer inspection  reveals small glitches.
In the figure, the statistical error bars are smaller than the symbol size and the red and blue dots refer to the two series of measurements.
By extracting the derivative of the energy-weighted excitation function~\cite{Esbeosc} D(E) we obtain rather regular oscillations as shown in Fig.~\ref{fig2}, although the uncertainties become obviously larger. 
\par
It is worth noting that 
two well defined peaks at E$_{c.m.}\simeq$35 and 36.5 MeV definitely show up in both series of  data (I and II run in the figure). 
This reassures us about  the quality of the performed measurements.
 However, a third peak at E$_{c.m.}\simeq$33.5-34 MeV is clearly observed only in the II run, due to 2-3 points of the I run being significantly lower. In our experience, this originates from very small unwanted changes of the beam conditions (focusing, direction) when changing the energy,  beyond the controls 
of the diagnostics. The consequences are hardly visible in the excitation function (see Fig.~\ref{fig1}(a)), but the representation in terms of its $\sl derivative$  amplifies the effect.

\section{Coupled-channels analysis}

 CC calculations have been performed, using the same formalism and structure input recently employed for the low- and near-barrier data of $^{28}$Si +$^{28}$Si~\cite{2828_2014}. In the following, as in that  work, the Ch1 and Ch10 calculations refer to the no-coupling limit and to the 
calculation where 10 coupled channels have been considered.
 The M3Y+repulsion potential~\cite{Misicu} has been used, as discussed below.
\par
The differences from the potential used for the sub-barrier data~\cite{2828_2014} are
a larger diffuseness of the imaginary potential, $a_w$ = 0.3 fm instead of $a_w$ = 0.2 fm, and 
a slightly smaller density radius of $^{28}$Si, $R=$ 3.125 fm instead of $R=$ 3.135 fm.
These modifications were made because they improve the fit of the Ch10
 calculation to the  data of Ref.~\cite{2828_2014} by reducing the $\chi^2/N$ from 1.7 to 1.2.
The diffuseness of the density that is used to construct the
M3Y potential, as well as the diffuseness associated with the 
repulsion, and the strength of the imaginary potential,
have not been varied ($a$ = 0.48 fm $a_r$ = 0.398 fm and $W_0$ = 5 MeV, respectively).

\par
The calculated excitation function is compared to the data in Fig.~\ref{fig1}(b). 
The corresponding derivative D(E) is shown in Fig.~\ref{fig2} (red curve) and is in fairly good 
agreement with the observed oscillations.
We point out that the present Ch10 calculation was calibrated to fit   the measured low-energy cross sections of Ref.~\cite{2828_2014}.
The prediction it makes at higher energies agrees well  with
  the new data shown in Fig.~\ref{fig1}(b) and the first derivative of $E\sigma$
  shown in Fig.~\ref{fig2} is also in fairly good agreement with the data.
It is remarkable that the whole set of data, including the oscillations, are now reproduced within a single theoretical model.

The old data of Ref.~\cite{Gary} (not shown here) were originally analyzed by the CC calculations of Ref.~\cite{Esbeosc}
 that used a rather large radius parameter R = 3.17 fm, and a$_r$= 0.378 fm, with a relatively strong imaginary potential ($a_w$ = 0.5 fm, $W_0$ = 10 MeV).
The blue line in Fig.~\ref{fig2} is that calculation which agrees with the present observations rather well.
However, that calculation does not fit the recent sub-barrier cross sections~\cite{2828_2014} (Fig.~\ref{fig1}(b), blue curve) at all, even when a weaker imaginary potential is used. This leads us to the important conclusion that
measuring the sub-barrier excitation function is essential for disentangling the ambiguities in the choice of the ion-ion potential that arise when only considering the data above the barrier.
\par
The result of the no-coupling calculation Ch1 using the potential of this work is also reported in Fig.~\ref{fig2} as a black dashed line. Weak oscillations can be observed  providing a poor fit to the experimental data.
A strong effect of couplings on the position and amplitude of the peaks is obviously seen.
It is tempting to associate the observed oscillations to the penetration of successive centrifugal barriers, as in lighter systems
where neighboring barriers are well separated. 
However, this association turns out to be somewhat
distorted as discussed below.
 \begin{figure*}[ht!]
\includegraphics[width=8.5cm]{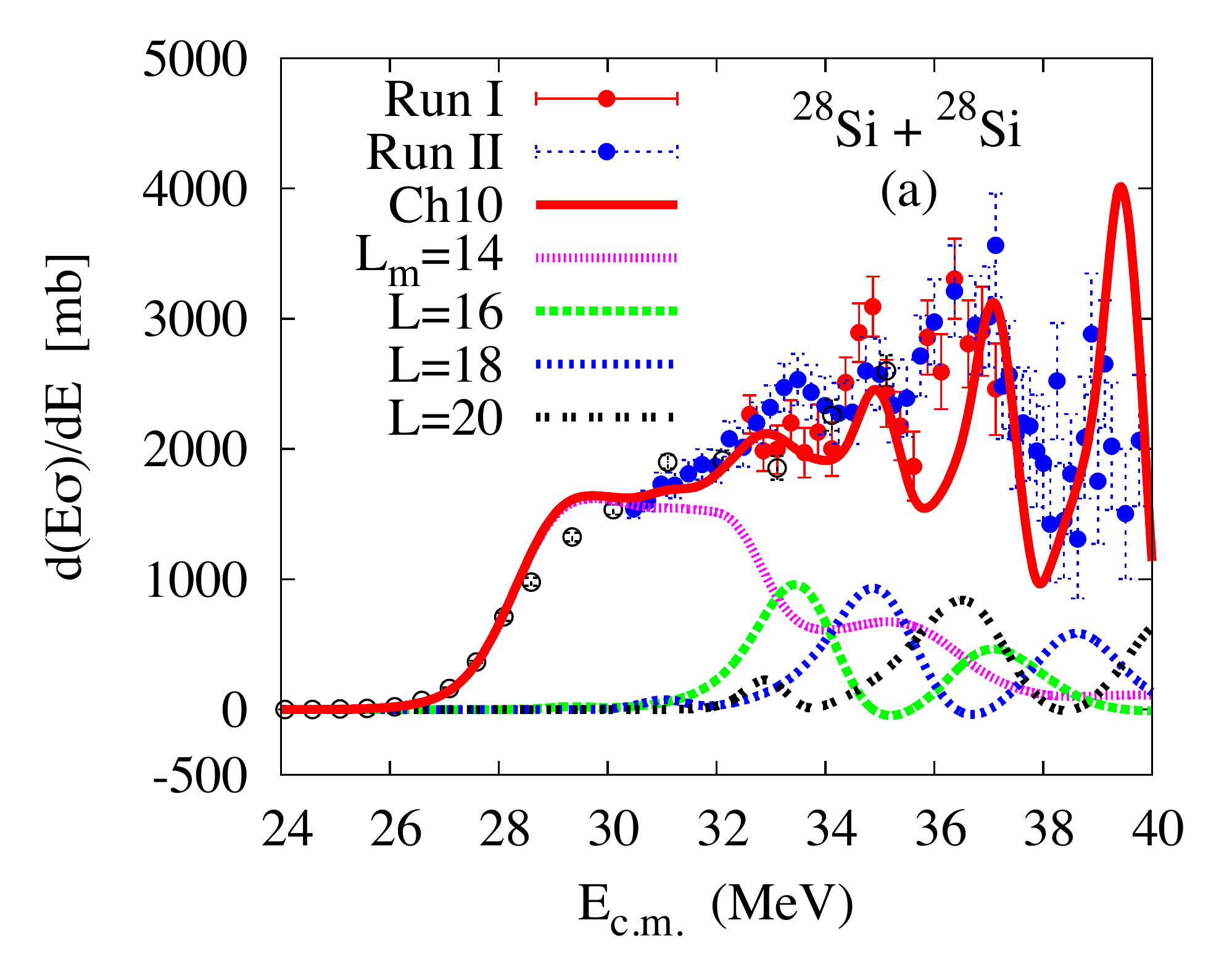}
\includegraphics[width=8.5cm]{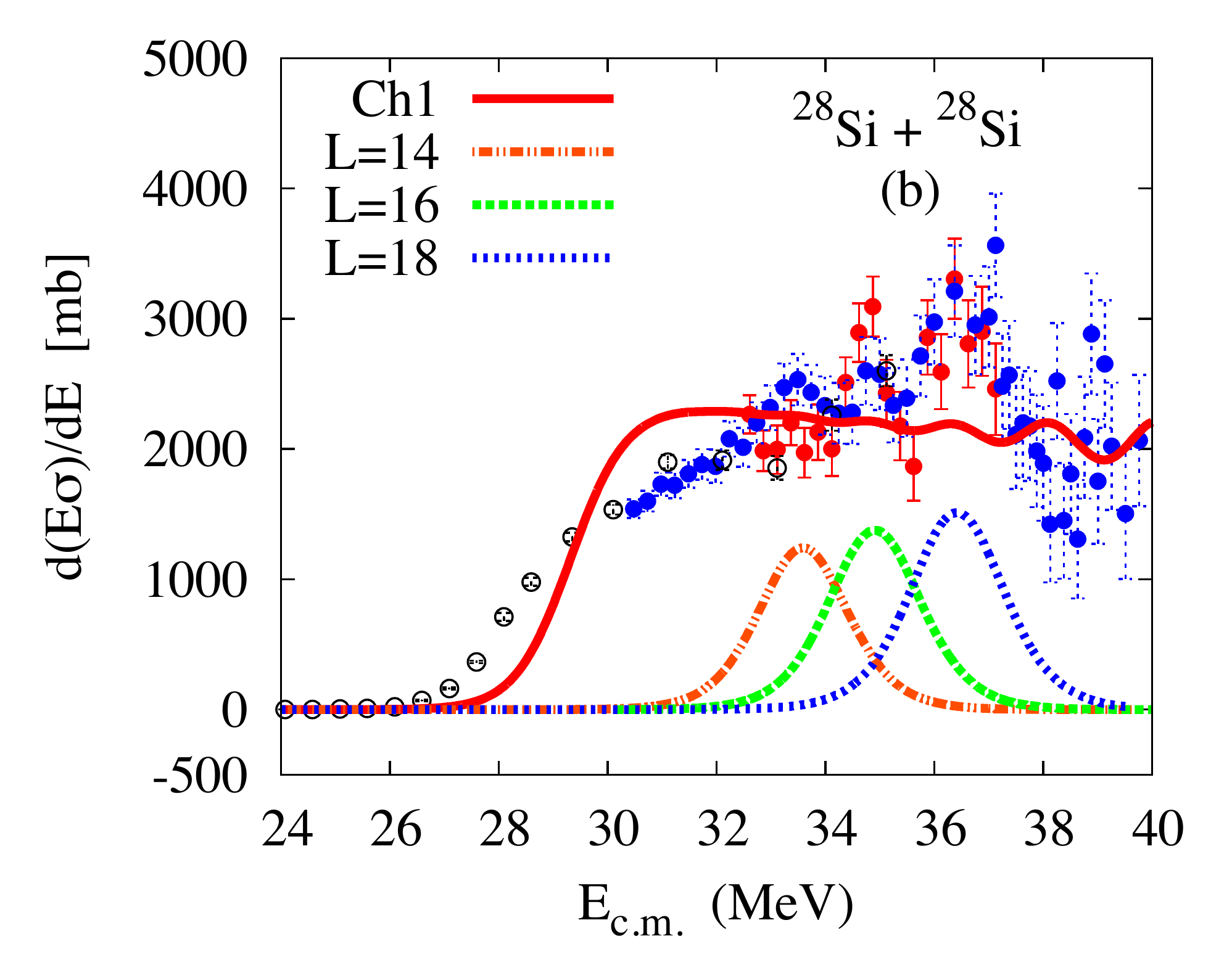}
\caption{(color online) (a) The first derivative of the energy-weighted 
cross sections. The results of Ch10 calculations are shown.
The solid red curve is the full result with a maximum angular
momentum of $L_{m}$ = 38. The result for $L_{m}$=14 is also shown and
so are the individual contributions for $L$ = 16-20. 
(b) The results of Ch1 calculations are shown. The experimental peaks at 33, 35, and 37 MeV are associated with $L$ = 14, 16, and 18, while they are primarily associated with $L$ = 16, 18,
and 20 in Ch10 calculations.}
\label{fig3}
\end{figure*}

Fig.~\ref{fig3}(a) shows again the observed oscillations together with the  results of the Ch10 calculation.
The individual contributions to $D(E)$ from the angular momenta $L$ = 16-20 
are also plotted. 
The comparison shows that the experimental peaks near 
33, 35 and 37 MeV correlate with the peaks produced by the angular 
momenta $L$ =16, 18, and 20. However, the height of these peaks 
constitute less that 50\% of the total distribution, partly because 
the peaks overlap and partly because the peaks are fragmented as 
discussed below. It is therefore concluded that it is not possible 
to assign a particular angular momentum to each experimental peak. 
Rather, each peak receives  contributions from several $L$-values.
 For example, the small peak that appears near 33 MeV for $L$ = 20
    is believed to be a real and not a spurious peak because a similar
    small peak appears for the other values of $L$. However, they are not
    visible in the linear plot of Fig.~\ref{fig3}(a).
\par
Ch1 calculations, i.~e., without any coupled-channels effects (see also Fig.~\ref{fig2}),
are reported in Fig.~\ref{fig3}(b). The Ch1 calculation does not contain any strong peaks.
This is because the peaks for $L$ = 14, 
16, and 18 in the no-coupling limit are broad and overlap, so that  their sum is essentially flat.
It is also seen that the three experimental peaks correlate with the three
calculated peaks for $L$ = 14, 16, and 18. 
This correlation, nevertheless, does not allow one to assign an angular
momentum to the experimental peaks, 
because the strong coupled-channels effects  lower and fragment the 
effective centrifugal barriers. 

The present data, along with the Ch1 and Ch10 calculations, point to the essential contribution of 
channel couplings for the appearance of oscillations. The magnitude of the oscillations is quite sensitive to the
   strength and the diffuseness of the imaginary potential.
   However, the peak positions are insensitive to these parameters.
   The diffuseness  $a_w$ = 0.3 fm was chosen because it
   optimises the fit to the old data of Ref.~\cite{2828_2014}.
   It produces oscillations in Ch10 calculations that are in
   reasonable agreement with the new data, which demonstrates
   a consistency of the old and the new data sets.
This is valid for $^{28}$Si + $^{28}$Si, but it is not a general conclusion. Indeed, in other (even near-by) systems, coupling effects might as well destroy oscillating structures reminiscent of penetration of successive $L$-barriers.


   We point out that the shallow potential we have used  produces rather thick centrifugal
   barriers and the associated penetration factor will therefore change quickly
   as the beam energy increases across the barrier height. The rapid rise
   of the penetration factor results in relatively narrow peaks as illustrated  in Fig.~\ref{fig3}(a).
   
\begin{figure*}[ht!]
\includegraphics[width=8.5cm]{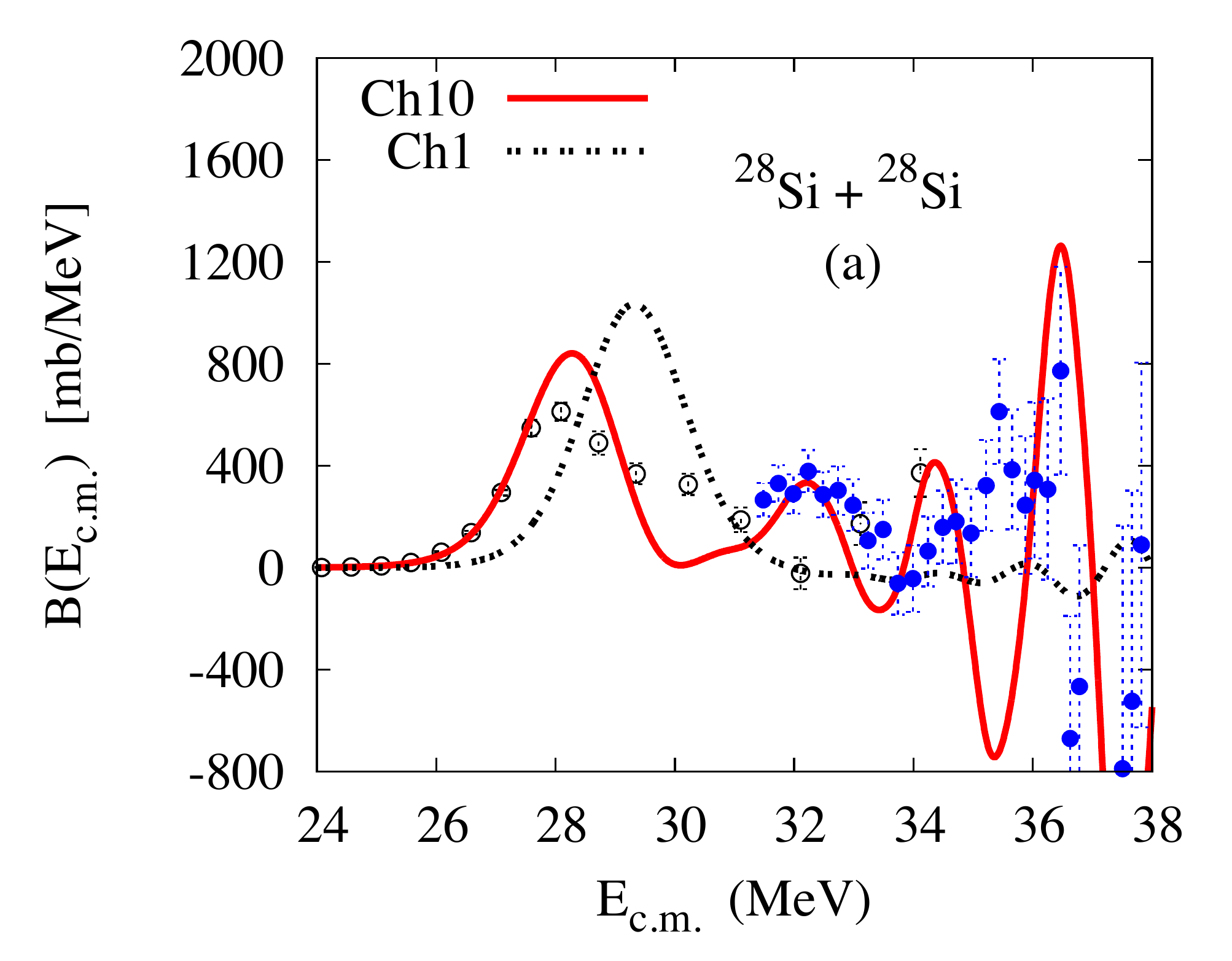}
\includegraphics[width=8.5cm]{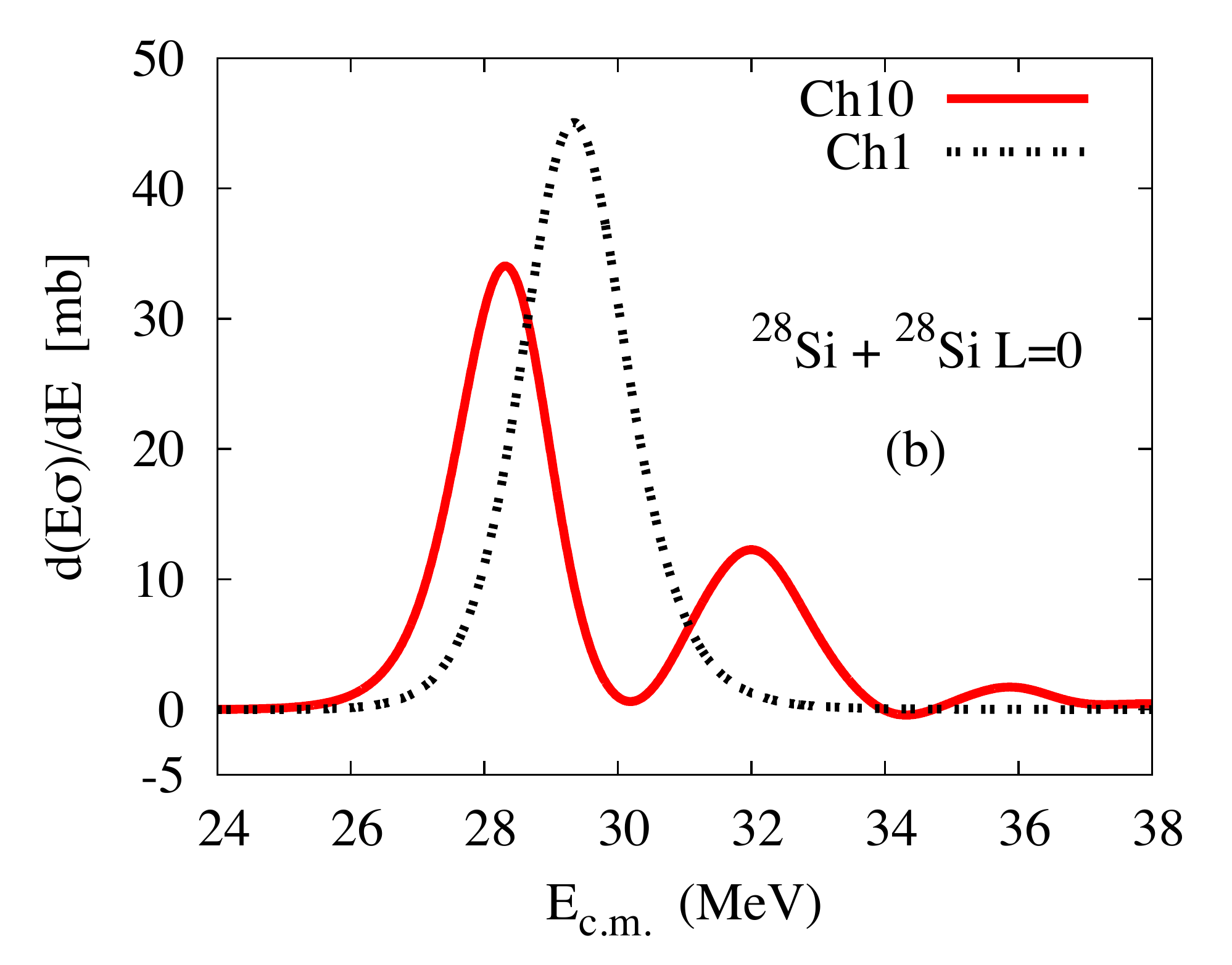}
\caption{(color online) Barrier distributions defined in Eq.~(\ref{cent})
and obtained from the present Run II data 
and from Ref.~\cite{2828_2014} are 
compared in (a) to distributions obtained from the Ch1 and Ch10 calculations discussed in 
the text. 
Panel (b) shows the first derivative of the energy weighted cross 
sections calculated for the angular momentum $L$=0.}
\label{fig4}
\end{figure*}

\section {Fusion barrier distributions}

In general, the barrier distribution one obtains as the second derivative of the energy-weighted experimental cross sections 
or of CC calculations (see Eq.~(\ref{cent})) may contain several peaks, due to couplings to other reaction channels during the fusion process.
The distributions extracted from the Ch1 
and Ch10 calculations discussed in this paper using the M3Y+rep interaction, are compared in
Fig.~\ref{fig4}(a) to the results of the previous measurement~\cite{2828_2014} 
and  Run II of the present measurement. 
\par
The near- and sub-barrier data produce a strongly asymmetric distribution that peaks at 28 MeV.  The present data reveal a second peak at 32 MeV, in good agreement with the Ch10 calculation, although the first peak is somewhat over-predicted while the second one is well accounted for.
\par
In this respect it is useful to plot (see Fig.~\ref{fig4}(b)) the first derivative of the 
energy-weighted cross section for the angular momentum $L$= 0 which is peaked at the Coulomb barrier in the Hill-Wheeler approximation~\cite{HillWheeler}.  
It resembles the barrier distribution shown in Fig.~\ref{fig4}(a), both with respect to the first and the second peak obtained in the Ch10 calculation. As discussed in the {\it Introduction}, this is a nice confirmation of the original idea of Rowley et al.~\cite{rowley},
namely, that the second derivative defined in Eq.~(\ref{cent})  can be interpreted as a barrier distribution. In particular, the experimental peak observed at 32 MeV in Fig.~\ref{fig4}(a) is a real peak of the barrier distribution because it is confirmed by both of the Ch10 calculations that are shown in Fig.~\ref{fig4}(a) and Fig.~\ref{fig4}(b), respectively.

\par
The experimental barrier distribution 
shown in Fig.~\ref{fig4}(a) develops some structures at energies above 
the peak at 32 MeV that do not appear convincingly in Fig.~\ref{fig4}(b).
In that energy range, we have evidence that the contributions of successive $L$-values gradually lose overlap. This implies the breakdown of  Wong's formula, and, consequently, of the interpretation of the second derivative of Eq.~(\ref{cent}) 
as a barrier distribution for a system as heavy as $^{28}$Si + $^{28}$Si.

\section {Conclusions}

In conclusion, we have shown the results of detailed measurements of the fusion excitation function of $^{28}$Si + $^{28}$Si with very small energy steps, which reveal regular oscillations, best evidenced in the first derivative of the energy-weighted  excitation function. We have been able to reproduce these high-energy oscillations and the sub-barrier cross sections,  within the same coupled-channels model using the shallow M3Y+repulsion potential. 

It appears that the existence of oscillations is tightly bound to channel couplings in this relatively heavy system, while in lighter cases  the oscillations have been suggested to be  related to the overcoming of successive centrifugal barriers well spaced in energy. 
In $^{28}$Si + $^{28}$Si, the oscillations do appear, 
but the one-to-one relation  between each peak and the height of a centrifugal barrier is lost because of strong coupling effects. 
Checking the importance of the oblate deformation of $^{28}$Si in this, calls for an analogous  experiment  on the nearby system $^{30}$Si + $^{30}$Si 
because $^{30}$Si is essentially a spherical nucleus.

As the last point of this article we also suggest that, for $^{28}$Si + $^{28}$Si,  the interpretation of the second derivative of the excitation function as a barrier distribution breaks down at energies well above the Coulomb barrier.

\section{Acknowledgements}
We acknowledge the highly professional work of the XTU Tandem staff during the beam times, and of M.Loriggiola for excellent target preparation.
The research leading to these results has received funding from the European Union Seventh Framework
Programme FP7/2007- 2013 under Grant Agreement No. 262010 - ENSAR. This work has been supported in part by  Croatian Science Foundation under the project 7194. H.E. is supported by the US Department of Energy, Office of Science, Office of Nuclear Physics, Contract No. DE-AC02-06CH11357.

\end{document}